\documentclass[twocolumn,prl,showpacs,superscriptaddress]{revtex4}
\usepackage[latin9]{inputenc}
\usepackage{graphicx}

\makeatletter
\@ifundefined{textcolor}{}
{%
 \definecolor{BLACK}{gray}{0}
 \definecolor{WHITE}{gray}{1}
 \definecolor{RED}{rgb}{1,0,0}
 \definecolor{GREEN}{rgb}{0,1,0}
 \definecolor{BLUE}{rgb}{0,0,1}
 \definecolor{CYAN}{cmyk}{1,0,0,0}
 \definecolor{MAGENTA}{cmyk}{0,1,0,0}
 \definecolor{YELLOW}{cmyk}{0,0,1,0}
}

\usepackage{dcolumn}\usepackage{bm}

\makeatother

\begin{document}

\title{Reply to Comment on ``State-independent experimental test of quantum
contextuality in an indivisible system''}

\maketitle
The comment by Amselem et al. \cite{1} misinterprates the logic and
assumption of our experiment \cite{2}. Note that for tests of quantum
contextuality, so far no experiment can be done in a loophole-free
and device-independent manner. We need to make some reasonable assumptions
in experiments to rule out the noncontexual hidden variable models.
What we have assumed in our experiment is about functioning of some
simple linear optical devices: half wave plates (HWP) and polarization
beam splitters (PBS). Basically, we assume that a HWP, set at an angle
$\theta$, transforms the polarization $H,V$ of the incoming light
field by $H\rightarrow cos(2\theta)H+sin(2\theta)V$, $V\rightarrow-sin(2\theta)H+cos(2\theta)V$
and a PBS transmits the light component in $H$-polarizaiton and reflects
its component in $V$-polarization \cite{2}. This knowledge does
not require assumption of fomalism of quantum mechanics and can be
regarded as basic experimental facts/laws about these well-calibrated
linear optical devices. The linear transfomormation of these optical
modes is apparently independent of the intensity of the incoming light
and holds in classical optics as well as in quantum case.

A schematic setup of our experiment is shown in Fig. 1. The mode tranformer
composed of the PBS and the HWPs link the modes $A_{i},A_{j},A_{k}$ right before
the light detectors with the modes $0,1,2$, which are prepared in
the same state for different experimental trials. The light detector
behaves like a black box, which gives binary measurement outcomes
(click or no-click) for the incoming field/mode. We assume the detectors
are identical and exchangable as it is the case in experiments. For
test of contextuality, we just need to make sure that the observable
$A_{i}$ before the detector $D_{i}$, expressed in term of the modes
$0,1,2$, remains the same when we change the observable $A_{j}$
to $A_{j'}$ before the other detector for measurement of the correlations
\cite{3}. With knowledge of functioning of the HWPs and the PBS in the
mode transformer, one can easily check that this is the case in our
experiment when we tune the angles of the HWPs. For some trials of
the experiment, we swap the labeling of the modes $2$ and $0$(1).
Again, with knowledge of functioning of the HWPs and PBS, we are still
measuring the same observable, which, expressed in term of the relabeled
modes $0,1,2$,, is under the same system state.
\begin{figure}
\includegraphics[width=8.5cm]{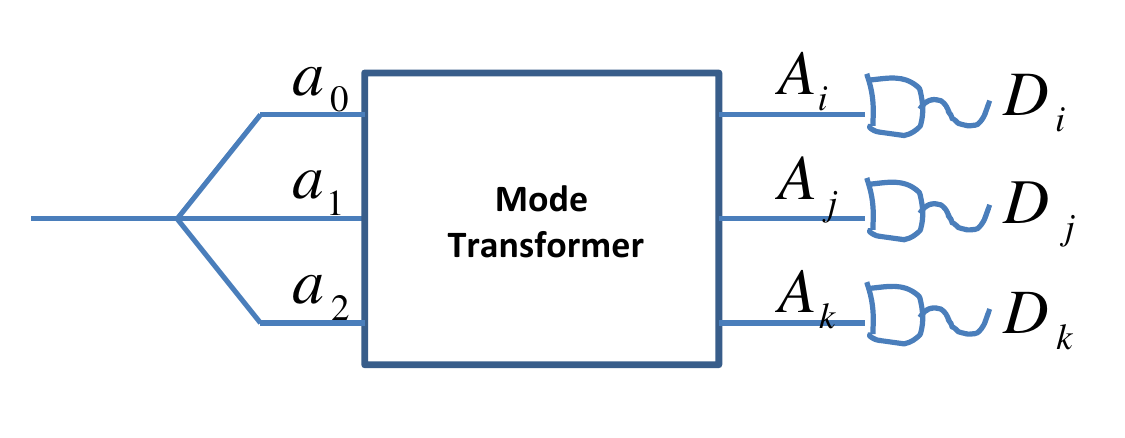}

\caption{Illustration of the schematic experimental setup. }
\end{figure}

Note that the functioning of these linear optical devices are also
assumed in previous experiments on test of quantum contextuality.
For instance, in Ref. \cite{4}, the real experimental setup is shown
in Fig. 3 there. To reduce the real setup to the schematic setup shown
in Fig. 1 there for test of quantum contextuality, one has to assume
that the PBS and the HWPs set at the right angles transform the optical
modes as they are supposed to function. So this assumption is not
particular to our experiment at all.

This work was supported by the National Basic
Research Program of China (973 Program) 2011CBA00300 (2011CBA00302) and the
NSFC Grant 61033001.

C. Zu$^{1}$, Y.-X. Wang$^{1}$, D.-L. Deng$^{1,2}$, X.-Y. Chang$^{1}$,
K. Liu$^{1}$, P.-Y. Hou$^{1}$, H.-X. Yang$^{1}$, L.-M. Duan$^{1,2}$

$^{1}$Center for Quantum Information, IIIS, Tsinghua University,
Beijing, China;$^{2}$Department of Physics, University of Michigan,
Ann Arbor, Michigan 48109, USA


\begin{thebibliography}{1}
\bibitem{1} E. Amselem et al., the proceeding comment.

\bibitem{2} C. Zu et al., Phys. Rev. Lett. 109, 150401 (2012).

\bibitem{3} S. X. Yu and C. H. Oh, Phys. Rev. Lett. 108, 030402-1-5
(2012).

\bibitem{4} R. Lapkiewicz et al., Nature (London) 474, 490 (2011). \end{thebibliography}
\end{document}